\documentclass[aps,pre,onecolumn,10pt,superscriptaddress,showpacs,amsmath,amssymb]{revtex4-1}

\usepackage{hyperref}
\usepackage{color}
\usepackage{graphicx}
\newcommand{\ave}[1]{\langle #1\rangle}
\let\d\relax
\DeclareMathOperator{\d}{d}

\begin{document}

	\title{Interplay of network dynamics and ties heterogeneity on spreading
	dynamics}
	
	\author{Luca Ferreri}
	\email{luca.ferreri@unito.it}
	\affiliation{GECO-Computational Epidemiology Group, Department of
	Veterinary Sciences, University of Torino, largo Braccini 2, IT-10095 Grugliasco (TO)}
	\affiliation{ARCS - Applied Research on Computational Complex Systems
	Group, Department of Computer Science, University of Torino, corso Svizzera 185, IT-10149 Torino}
	\author{Paolo Bajardi}
	\affiliation{GECO-Computational Epidemiology Group, Department of
	Veterinary Sciences, University of Torino, largo Braccini 2, IT-10095 Grugliasco (TO)}
	\affiliation{ARCS - Applied Research on Computational Complex Systems
	Group, Department of Computer Science, University of Torino, corso Svizzera 185, IT-10149 Torino}
	\author{Mario Giacobini}
	\affiliation{GECO-Computational Epidemiology Group, Department of
	Veterinary Sciences, University of Torino, largo Braccini 2, IT-10095 Grugliasco (TO)}
	\affiliation{ARCS - Applied Research on Computational Complex Systems
	Group, Department of Computer Science, University of Torino, corso Svizzera 185, IT-10149 Torino}
	\affiliation{CSU-Complex Systems Unit, Molecular Biotechnology Centre,
	University of Torino, via Nizza 52, IT-10126 Torino}

	\author{Silvia Perazzo}
	\affiliation{Department of Mathematics ``Giuseppe Peano'', University of
	Torino, via Carlo Alberto 10, IT-10123 Torino}
	
	\author{Ezio Venturino}
	\affiliation{Department of Mathematics ``Giuseppe Peano'', University of
	Torino, via Carlo Alberto 10, IT-10123 Torino}

	\date{\today}
	\begin{abstract}
	The structure of a network dramatically affects the spreading phenomena unfolding
	upon it. The contact distribution of the nodes has long been
	recognized as the key ingredient in influencing the outbreak
	events. However, limited knowledge is currently {available} on the role of
	the weight of the edges on the persistence of a pathogen. At the same time, recent
	works showed a strong influence of temporal network dynamics on disease
	spreading. In this work we provide an analytical understanding,
	corroborated by numerical simulations, about the conditions for infected
	stable state {in} weighted networks. In particular, we reveal the role of
	heterogeneity of edge weights and of the dynamic assignment of weights
	on  the ties in the network in driving the spread of the epidemic. In this context we show that when weights are dynamically assigned to ties in the network an heterogeneous distribution is able to hamper the diffusion of the disease, contrary to what happens when weights are fixed in time.
	\end{abstract}
	\pacs{89.75.Hc
	, 05.45.Xt
	, 87.23.Ge
	}
	\maketitle
	
	\section{Introduction}
	In the last decade, an increasing scientific effort has been devoted to the
	{understanding and characterization} of spreading phenomena in complex settings,
	ranging from computer viruses to rumors, human diseases and animals
	pathogens. To this aim, the study of diffusion processes on complex
	networks represents a major contribution to move beyond the homogeneous
	mixing approximation and to investigate the effect of the interaction of hosts
	in shaping the epidemic spreading. { Heterogeneous networks are usually described as annealed or quenched. While an annealed network is defined by an adjacency matrix fixed only on average \cite{Dorogovtsev2008}, the latter is defined by a fixed adjacency matrix. In this context, until today, there have been two main analytical approaches that could be exploited to describe dynamical process occurring on heterogeneous networks: the heterogeneous mean-field (HMF) theory and the spectral approach \cite{Castellano2006,Castellano2010}. More specifically, the HMF approach is suitable for annealed networks \cite{Dorogovtsev2008} and predicts an epidemic threshold inversely proportional to the second moment of the network's degree distribution. In addition, the HMF approach is also able to capture the epidemic threshold for some classes of quenched networks (QN), for instance when the degree is power-law distributed with exponent between $2$ and $2.5$, \cite{Castellano2010}. However, it has been demonstrated that the phase transition of QN is properly described by the spectral radius of the adjacency matrix, \cite{Wang2003,Chung2003,Prakash2010,Gomez2010}. Both} 
	approaches pointed out the dramatic effects of contact heterogeneities
	in lowering the epidemic threshold
	\cite{lloyd2001,Pastor-Satorras2001,Bansal,Castellano2010,Barrat2008}.
	Beyond the degree-centrality, other measures have been recognized to be
	important indicators of the role of a node in the diffusion process, such as
	the betweenness centrality, \cite{Freeman1978,Friedkin1991}, closeness,
	\cite{Sabidussi1966,Dangalchev2006}, eigenvector and PageRank,
	\cite{Bonacich1972,Brin1998}, and k-shell
	\cite{Kitsak2010,Garas2012,Castellano2012}.
	
	At the same time, great interest has arisen in the last years in understanding and exploring evolving graphs
	\cite{Holme2012,Holme2013,Mukherjee2013,Pfitzner2013,Rocha2013b} and
	their interplay with dynamical process occurring on them. Schwarzkopf
	and colleagues, \cite{Schwarzkopf2010}, for instance, introduced and
	analyzed a model for epidemic spreading on a rewiring network,
	highlighting that the rewiring process tends to suppress the infection.
	Similarly, Volz and his collaborators,
	\cite{Volz2007,Volz2009,Miller2012} worked on the impact of the contact
	duration on the pathogen's diffusion. Meanwhile, other models have been
	proposed describing the coevolution of the graph and the spreading
	process,
	\cite{Gross2006,Gross2008,VanSegbroeck2010,Marceau2010,Taylor2012,Masuda2013},
	showing the non-linear interplay of the two dynamics. More recently,
	Perra and collaborators, \cite{Perra2012}, introduced an activity driven
	model in which the instantaneous interaction of agents is defined by the
	activity potential and provided an analytical description of the
	epidemic threshold in such context. In particular, via an HMF
	approximation, they showed that the stable infected state { depends on the first and the second moments of the nodes activity distribution probability.}
	
	While many researches have been conducted on unweighted graphs up today,
	a limited knowledge,
	\cite{Yan2005,Meloni2009,Baronchelli2010,Deijfen2011,Chu2011,Britton2011,Britton2012,Yang2012,Kamp2013,Miritello2011},
	is currently available on the effect of the weight of the links on pathogen
	spreading over an evolving graph. {This} is a relevant issue for exploring
	real-world scenarios, where dynamical contacts between hosts are observed
	\cite{Cattuto2010,Bajardi2012,Rocha2011,Rocha2013} and a weighted
	representation is often required
	\cite{Eames2009,Stehle2011,Bajardi2011,Machens2013,Miritello2013}. In this paper we
	are interested in understanding how the combination of heterogeneity of
	edge weights and the dynamic assignment of weights on network's ties
	could affect the epidemic spreading. From an epidemiological point of
	view, we can interpret weights on ties as an indicator of the intensity or
	duration of the interactions between two connected nodes representing the
	hosts of the pathogen. There exist several
	examples in real-world settings where a weighted representation of contacts between network
	nodes is needed \cite{Barrat2004}: from the early work on sociograms \cite{moreno}, where
	weights describe the number of real contacts among two people, to email
	networks, where weights represent the number of exchanged emails, to
	cattle trade movements, where weights take into account the number of
	animals moved between livestock premises \cite{Bajardi2011}. Although
	the intuition and common knowledge suggest that heavier weights should indicate greater ease of transmission, a comprehensive
	understanding of the system dynamics is still lacking. To bridge the gap
	between intuition and formal investigation, and to extend our
	findings to dynamical networks, we define a theoretical and manageable
	framework to perform both analytical calculations and computational
	experiments.
	
	\section{Graphs with stable weights pattern}
	\subsection{Graphs Generation}\label{grapgen}
	As a first step in analyzing the role of edge-weights in an epidemic
	spreading, we need a series of networks with some desired properties:
	an assigned degree-distribution, the same number of edges, an assigned
	weight heterogeneity, and the total amount of weights. To this end we
	work on graphs of size $N=10^4$. We initially create a sequence of stubs
	by pooling from a probability distribution $p(k)\sim k^{-2.25}$ (we
	choose as exponent of the power law $2.25$, a
	good compromise between induced variability and a sufficiently large average degree
	$\ave{k}\simeq5$), and then we tune the degree heterogeneity of the
	produced network by randomly reshuffling a fraction $\gamma_k\in [0,1]$
	of the stubs' origins. We would like to underline that when $\gamma_k=1$
	all the stubs' origins are reshuffled, resulting in a random assignment
	of contacts, while for $\gamma_k=0$ the network exhibits the maximum
	degree of heterogeneity as caused by exponent $2.25$. {Since the reshuffling is performed on the open stubs and not on the links, the reshuffling process destroys the original degree sequence of the nodes, thus defining the new network topology.}	
	
\begin{figure}
		\begin{center}
			\includegraphics{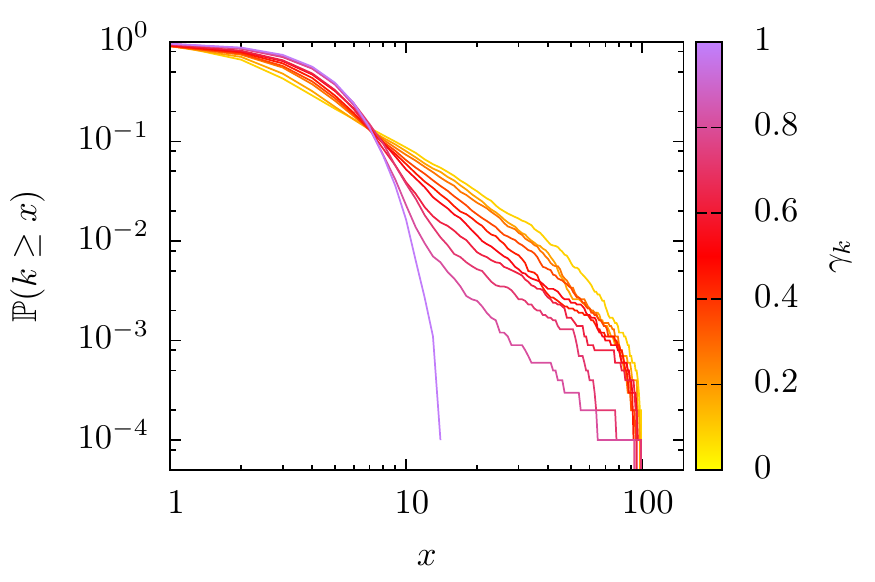}
		\end{center}\caption{ Degree distribution of networks varying $\gamma_k$ and maintaining $\gamma_w=0$. $\gamma_k$ ranges between $0$, yellow (upper) curve, to $1$, violet (lower) curve.}\label{distr}
	\end{figure}

\begin{figure}
		\begin{center}
			\includegraphics{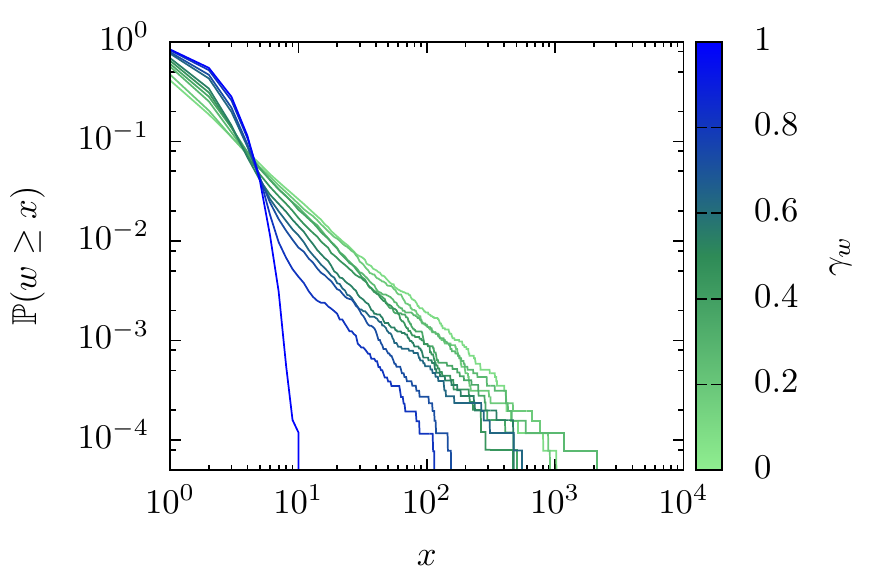}
		\end{center}\caption{ Weight distribution of networks varying $\gamma_w$ and maintaining $\gamma_k=0$. $\gamma_w$ ranges between $0$, light-green (upper) curve, to $1$, blue (lower) curve.}\label{distr2}
	\end{figure}

\begin{figure}
		\begin{center}
			\includegraphics{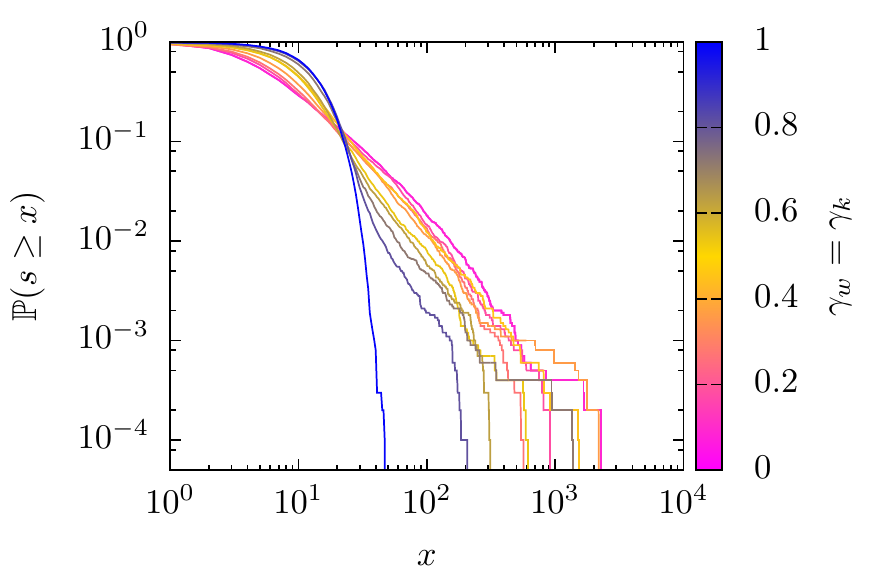}
		\end{center}\caption{ Strength distribution of networks varying $\gamma_w=\gamma_k$. $\gamma_w=\gamma_k$ ranges between $0$, magenta (upper) curve, to $1$, blue (lower) curve.}\label{distr3}
	\end{figure}

	 We further remind
	that a vanishing epidemic threshold is observed when the degree is power-law distributed with an exponent between $2$ and $3$. Therefore we
	considered an exponent of $2.25$ as our most heterogeneous case and, by
	increasing the proportion of rewired links, we systematically explored
	increasingly homogeneous systems. At the end of this procedure, we assigned a discrete weight to every
	stub by generating it from the probability
	distribution $q(w)\sim w^{-2.25}$. Again, we choose the exponent $2.25$
	because of its interplay between strong heterogeneities and not too small
	average edge-weight ($\ave{w}\simeq2.66$). In order to tune the
	heterogeneity of the weight distribution, similarly to the rewiring
	procedure described before, we randomly selected a fraction $\gamma_w$ of
	the allocated weights and randomly reassigned them so that the new
	$w'_{i\bullet}=w_{i\bullet} -1$ and $w'_{k\bullet}=w_{k\bullet} +1$,
	being $w_{i\bullet}$ the weight of a stub departing from node $i$. We
	further impose that stubs of weight $1$ could not be reallocated in
	order to avoid stubs with zero weights { and to keep the degree of the nodes constant}. When examining a routing system
	or an infrastructural network with packets or hosts moving between
	nodes, reallocating a fraction of the total weight across the network
	allows us to tune the heterogeneity of the weight distribution while keeping
	the total amount of traffic and average weights constant. Then, using
	the configuration model \cite{Molloy1995}, adjusted to generate
	uncorrelated networks \cite{Catanzaro2005}, we close the stubs by further
	imposing that a stub of weight $w$ is tied with a stub of the same
	weight {(i.e. being $w_{i\bullet}$ a stub departing  from $i$ and having weight $w$, it can be connected with any stub $w_{j\bullet}$ where $j\neq i$)}. With the described algorithm, we are able to generate networks
	with the same number of edges and total traffic and the desired degree
	and weight heterogeneities. We refer to Figure \ref{distr}-\ref{distr3}
	for some graph features obtained with various $\gamma_k$ and $\gamma_w$. {More in detail, we highlight that for $\gamma_k=0$ we let $k$ range between $2$ and $\sqrt{N}$, \cite{Molloy1995,Catanzaro2005}, while for $\gamma_w=0$ we let $w$ range between $1$ and $N$.}

	We further remark that in the following we are going to consider two classes of
	graph:
	\begin{description}
	\item[annealed] if the graph is defined only by the degree and weight sequences
	(i.e. the stubs are closed at each considered iterations)
	\item[quenched] if the graph is the result of the stubs
	closure (i.e. the weighted adjacency matrix is fixed in time).
	\end{description}
	\subsection{Disease Spreading}
	In the following section we focus on the Susceptible-Infectious-Susceptible
	(SIS) compartmental model \cite{Anderson&May}, where nodes are divided in
	two classes according to their health status. A susceptible node is
	infected by an infective neighbor with probability $\beta$ and an
	infective node recovers from the infection with probability $\mu$, thus
	becoming susceptible again.
	\subsubsection{Analytical description}
	In order to achieve a qualitative understanding on the critical behavior
	of the spreading phenomena on weighted networks, we use an HMF approach
	\cite{Barrat2008} that can deal with weighted transmission routes. The
	HMF approach is known to be exact for weighted annealed networks
	\cite{Baronchelli2010} while for quenched networks a spectral approach
	should be preferred. Using a strength block approximation
	(i.e. assuming that all nodes with the same strength are statistically
	and epidemiologically equivalent), we can describe the variation of
	the fraction of infectious nodes of strength $s$ as
		\[\frac{\d i_s}{\d t}=-\mu i_s+(1-i_s)\left[1-\left(1-\beta\right)^{s\Theta_s}\right],\]
	where $\Theta_s$ is the probability that an edge originating from an
	infectious node is connected to a susceptible $s$-node. Furthermore,
	assuming no strength correlation among connected nodes we have that
		\begin{equation}\Theta_s=\frac{1}{\ave{s}}\sum_{s'}s' P(s')i_{s'}=\Theta,\label{SIS_mod}\end{equation}
	where $P(s')$ is the probability for a node to have strength $s'$.
	Imposing the stability condition $\frac{\d i_s}{\d t}=0$, after some
	algebraic manipulations, we find
		\[\Theta=G(\Theta)=\frac{1}{\ave{s}}\sum_s sP(s)\frac{\left[1-(1-\beta)^{s\Theta}\right]}{\mu+\left[1-(1-\beta)^{s\Theta}\right]},\]
	which, at least to our knowledge, can not be analytically solved.
	However, in the considered domain, we have that $G(\Theta)\leq\Theta$
	and $G(0)=0$. Moreover, $G$ is an increasing and concave function. Thus,
	if $G'(0)>1$, there exists $\hat\Theta\in(0,1]$ such as
	$G\left(\hat\Theta\right)=\hat\Theta$. In conclusion, a non-zero solution
	exists if
		\[-\frac{\ln(1-\beta)}{\mu}>\frac{\ave{s}}{\ave{s^2}}.\]
	Now, we want to understand how the strength changes in function of the
	weight and degree distribution, respectively $q(w)$ and $p(k)$. To this
	end, the probability generating function (pgf), \cite{wilf}, of the
	degree is
		\[A(x)=\sum_k x^k p(k),\]
	while $B(x)=\sum_w x^w q(w)$ is the pgf of the weight probability
	distribution. Consider now a node connected with $k$ other nodes. We
	define $S_k=w_1+\ldots+w_k$ as the random variables of the strength of a
	node of degree $k$. The pgf of such variables is
		\[
			C(x,k)=\sum_zx^zP(S_k=z)=\sum_z x^zP(w_1+\ldots+w_k=z)=\left(\sum_h x^hq(h)\right)^k=\left(B(x)\right)^k.
		\]
	Now, if the degree of nodes is also random then the pgf of the strength
	of a node is
		\begin{multline*}
			D(x)=\sum_s x^sP(S=s)=\sum_s x^s \sum_h P(S_h=s)p(h)=\sum_hp(h)\sum_s x^sP(S_h=s)=\sum_hp(h)C(x,h)=\\\sum_h p(h)\left(B(x)\right)^h=A\left(B(x)\right).
		\end{multline*}
	From a pgf, $G(x)$, the $n$-th moment of the
	distribution can be calculated as follows: $\ave{k^n}=\left[\left(x\frac{\d }{\d
	x}\right)^nG(x)\right]_{x=1}$. Hence,
		\[\ave{s}=A'\left(B(x=1)\right)B'(x=1)=\ave{k}\ave{w},\]
	and
		\[
		\ave{s^2}=A''(B(1))\left(B'(1)\right)^2+A'(B(1))B''(1)+\ave{s}=\ave{k^2}\ave{w}^2-\ave{k}\ave{w}^2+\ave{k}\ave{w^2}.
		\]
	Therefore, we obtain the condition for pathogen persistence among
	the population:
		\[-\frac{\ln(1-\beta)}{\mu}>\left(\frac{\ave{k^2}\ave{w}}{\ave{k}}+\frac{\ave{w^2}-\ave{w}^2}{\ave{w}}\right)^{-1}.\]
	This implies that the infection probability, $\beta$, necessary for the
	pathogen to persist in the system, is a decreasing function of the
	heterogeneity of either of the weight or the degree distribution, or both.
			
	\subsubsection{Numerical simulations}
	The analytical results were further analyzed by performing numerical
	simulations of epidemic spreading on annealed networks
	generated by exploring $(\gamma_k,\gamma_w)$. For each
	$(\gamma_k,\gamma_w,\beta,\mu)$ we simulate $100$ epidemic scenarios
	unfolding on independently generated graphs, randomly choosing $10$
	infected nodes as the initial condition (we arbitrarily chose the number of
	$10$ infected nodes as an initial condition since we want to explore the
	infected stable state more than the invasion pattern and we want to
	avoid an exceptionally high number of scenarios with disease extinction)
	and observing the steady state at the endemic equilibrium from which we
	evaluate the fraction of infected nodes $\ave{i}$. { Since the equilibrium of a SIS model in the active phase is not unequivocally defined, we stated that an epidemic spreading reaches the equilibrium at time $t$ if the median of the prevalences $i(\tau)$ with $\tau\in[t-50,t]$ was within the $1\%$ percentile of the prevalences $i(\tau)$ with $\tau\in[t-100,t-50]$. In addition, it is also worth to stress that the equilibria (disease-free or not) were reached long before the maximum allowed time, $t_{\max}=10^4$.}

\begin{figure}[h]
			\begin{center}
				\includegraphics{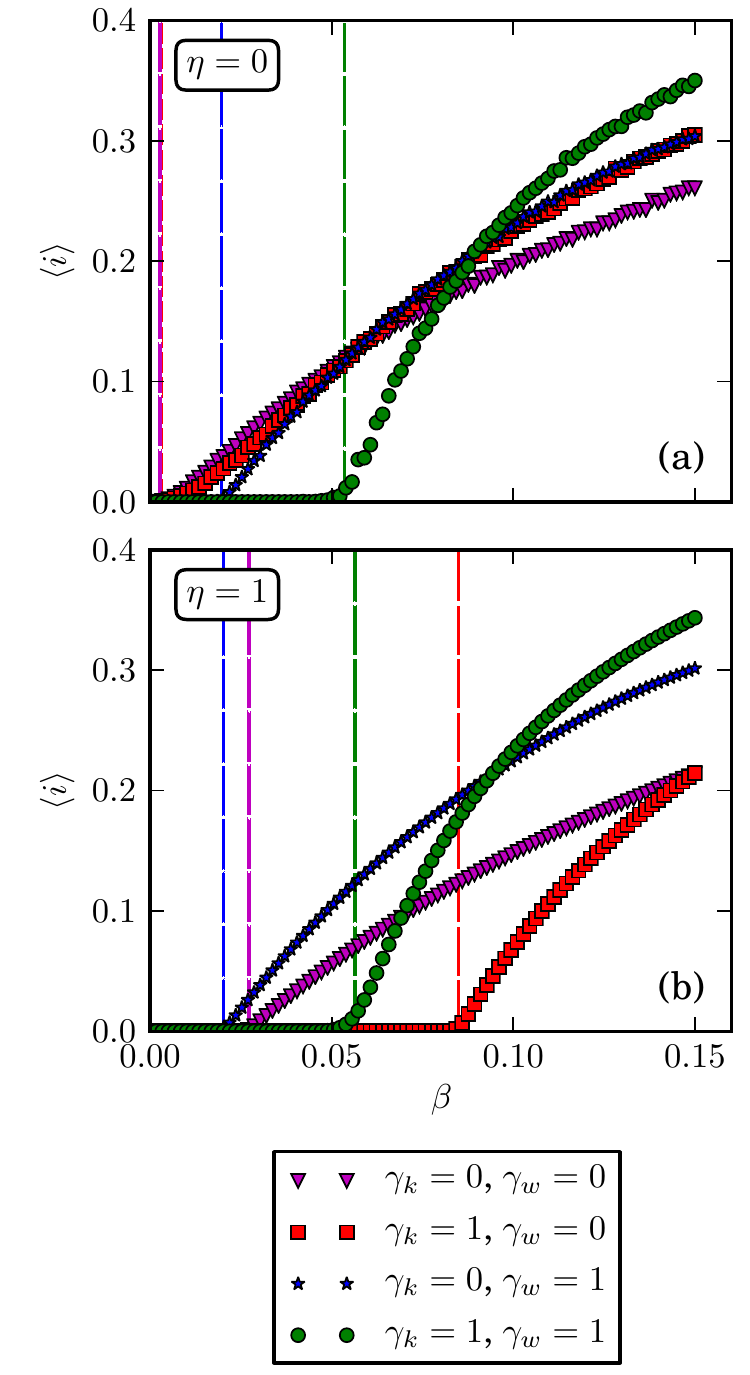}
			\end{center}
		\caption{ Fraction of infected nodes at endemic equilibrium
		reached as function of $\beta$, the transmission probability. For this figure $\mu$ is fixed to one.
		(a), results on annealed networks with stable weight pattern,
		$\eta=0$. (b), results on annelead networks with dynamical
		assignment of weights, $\eta=1$. Recovery probability $\mu$ is
		fixed to one. Vertical lines show the theoretical expectation of
		the thresholds. On graphs associated with $\eta=0$ the
		heterogeneity of both edge weights and node degrees seems to
		support pathogen persistence. On the other hand, on networks
		with dynamical assignment of weights, $\eta=1$, infection
		stability is more often achieved on graphs with
		homogeneously distributed weights.}\label{epi_curve}
		\end{figure}

	Numerical simulations
	reported at the top of Figure~\ref{epi_curve} show the influence of a
	broad strength distribution (i.e. $\gamma_k=0$ or $\gamma_w=0$) for the
	pathogen persistence. In particular, this picture shows the strong
	impact of the weight heterogeneity on the critical transmission
	parameter. Figure~\ref{epi_curve} also shows the good match between
	theoretical and empirical epidemic thresholds on annealed graphs. We
	underline that magenta and red lines seem almost coincident but 
	the red one is in fact slightly larger than the magenta, consistently with the
	observed epidemic curves. Empirical results on quenched networks show a
	similar patterns in terms of threshold effects for the different
	scenarios (data not shown) but, as expected, the HMF approximation does
	not provide a good estimation of the $\beta_c$.
	
\begin{figure}
			\includegraphics{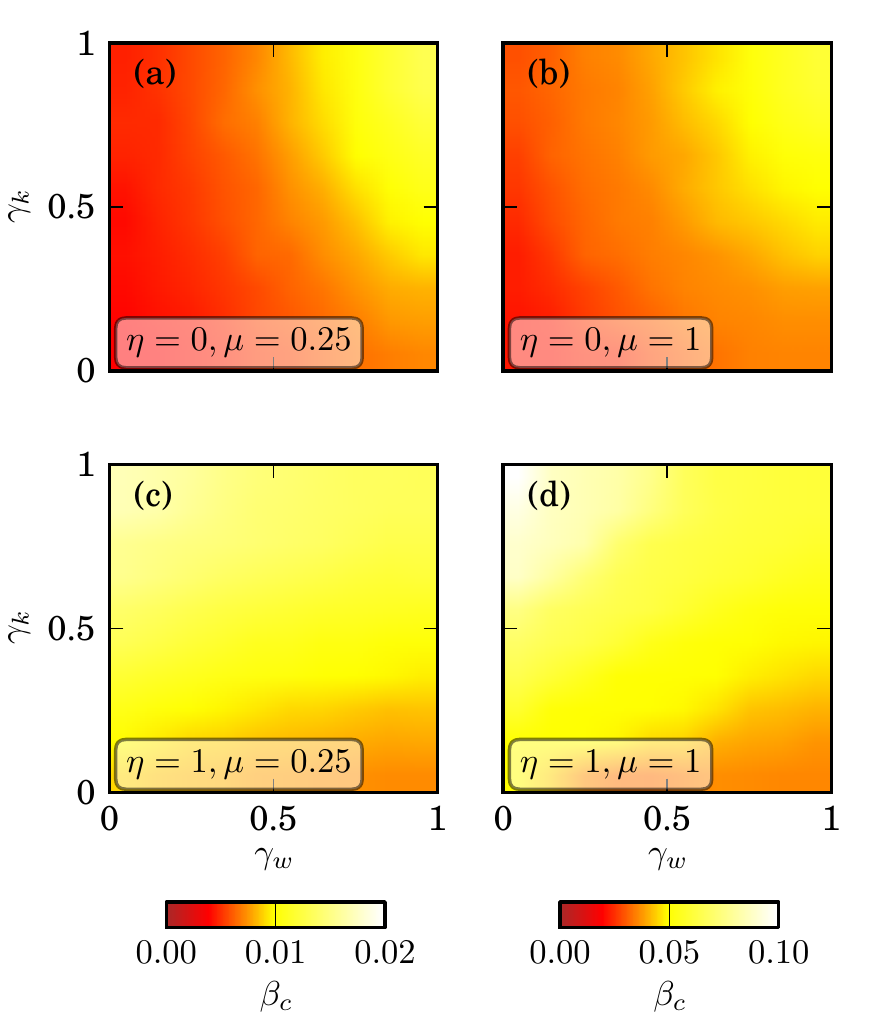}
		\caption{ Estimated epidemic threshold, $\beta_c$, as function of
		$\gamma_k$ and $\gamma_w$ on annealed networks. (a)-(c), recovery
		probability is $\mu=0.25$, (b)-(d), we fix $\mu=1$. (a)-(b), results obtained for
		graphs with stable weights pattern, $\eta=0$, (c)-(d), results
		obtained on graphs with dynamical assignment of weights,
		$\eta=1$. By comparing these two rows, a clear difference of
		behavior could be detected: for $\eta=1$ the edge weight
		heterogeneity ceases to be a favorable condition for infection
		spreading. We further explore this behavior in
		Fig.~\ref{pende}}\label{mat_bc_2}
		\end{figure}

	We integrate the aforementioned disease spreading with a hyperbolic
	variant of the Brent method for root-finding \cite{brent}. By using a
	tolerance $10^{-2}$ we detect the $\beta_c$ for which $\ave{i}$ is equal
	to $0.05$. $\ave{i}$ is measured as the averaged fraction of
	infected nodes at infection persistence in $100$ graphs. We explore $\mu=0.25$ and $\mu=1$ as the probability of recovery.
	Although the analytical description match fairly
	well with the numerical simulations, it is worth to emphasize that our
	interest lies in the qualitative investigation of $\beta_c$ for a wide
	range of networks with increasing heterogeneity rather than providing a
	quantitative estimate of the critical point. In Figure~\ref{mat_bc_2} we
	show the surface $\beta_c(\gamma_k, \gamma_w)$ obtained for annealed
	networks. Results confirm the well-known behavior
	\cite{Pastor-Satorras2001,Pastor-Satorras2001b,Barrat2008} of epidemic
	spreading on graphs: the larger the heterogeneity of connections among
	nodes, the lower the transmission probability needed for pathogen
	persistence. Moreover, in agreement with the analytical insights,
	the heterogeneity of weights also fosters pathogen persistence. We refer
	to the top picture of Figure~\ref{mat_bc_2} for an overview of these
	results.	
		
	
	\section{Dynamic assignment of weights}
	\subsection{Graph Generation}
\begin{figure}[h]
			\begin{center}
				\includegraphics{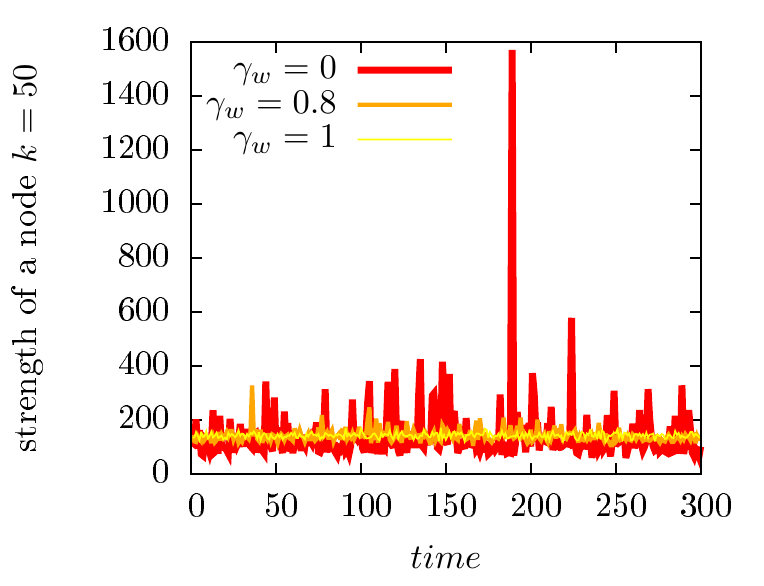}
			\end{center}
		\caption{ Strength of a randomly chosen node of degree $50$ in
		$(\gamma_k=0,\gamma_w=\{0,0.8,1\})$ networks with $\eta=1$
		fraction edge weights permuted at each time step}\label{spi}
		\end{figure}

\begin{figure}[h]
			\begin{center}
				\includegraphics{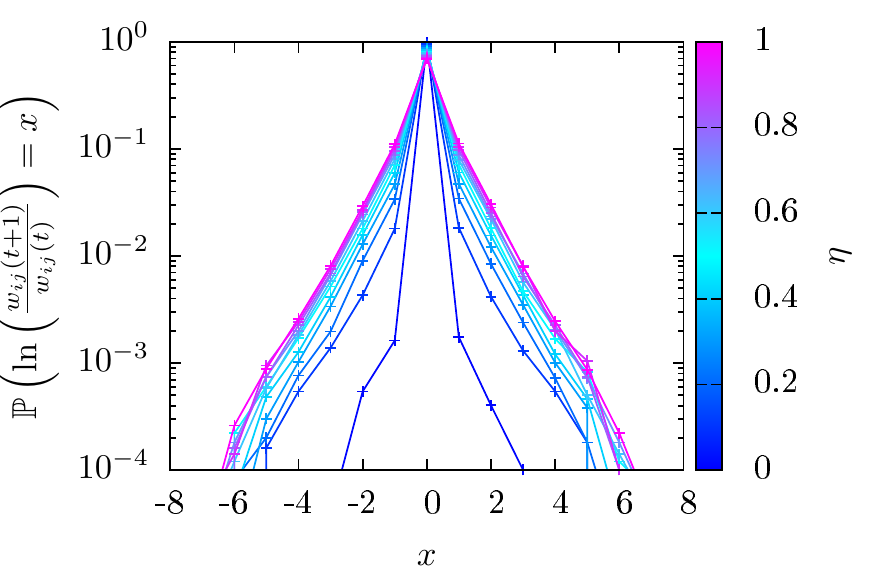}
			\end{center}
		\caption{ Probability density of the evolution rate of edge weights for
		dynamic heterogeneous networks $(\gamma_k=0,\gamma_w=0)$ for
		different $\eta$ values. $\eta$ ranges from $0$, blue (lower) curve, to $1$, magenta (upper) curve.}\label{rapp_temp}
		\end{figure}

	As a second step, we investigated how the dynamic assignment of weights on
	edges (i.e. when edges change their weight at each time-step) influences
	infection spreading. In particular, given a network (quenched or
	annealed) defined by a degree and weight sequence originated by couple
	$(\gamma_k,\gamma_w)$ as in Section~\ref{grapgen}, at each time step we
	permute a fraction $\eta\in [0,1]$ of weights. To provide the reader
	additional insights about the effect of this procedure, we plot in
	Figure~\ref{spi} the strength of a prototypical medium-highly connected
	node of degree $50$ in a $(\gamma_k=0,\gamma_w=\{0,0.8,1\})$ networks
	with $\eta=1$. In a $300$-step large time-window sudden spikes in the node's strength can be observed and these are caused by the re-assignation
	of weights on edges.
	We further explore the fluctuations of link properties as a consequence
	of the weight shuffling.  We consider the evolution rates of edge weights
	\[\ln\left(\frac{w_{ij}(t+1)}{w_{ij}(t)}\right)\]
	for different values of $\eta$. The distributions of those evolution
	rates are shown in Figure~\ref{rapp_temp} {and it is worth noticing that they resemble many real world systems \cite{stanley1996,Gautreau2009,Bajardi2011} where the evolution rate distribution shows an exponentially decaying tail.} Not surprisingly, the
	probability of a high evolution rate increases based on the
	fraction of edge weights reallocated, $\eta$. It is
	interesting to note that even for a small fraction of permuted weights
	(e.g. $\eta=0.3$) the weight of an edge could dramatically change, thus
	generating a bursted behavior. 

	\subsection{Disease Spreading}
	\subsubsection{Analytical description for $\eta=1$}
	Considering the complete dynamical case, i.e. $\eta=1$, we provide
	analytical insights about the epidemic threshold on annealed graphs. At
	each time step the weights of edges are given according to $q(w)$, i.e.
	the weight probability distribution, and with $\gamma_w$, i.e. the
	parameter responsible for tuning the heterogeneity of the distribution.
	Furthermore, assuming no correlation between the degree of nodes and the edge
	weights connecting them, and since the weights on edges are allocated at
	each time-step, we could distinguish nodes according to their degree,
	i.e. using degree block approximation. Therefore, the probability that a
	susceptible node connected with an infective one will be infected is
		\[\mathbb{P}=1-\left[\sum_w(1-\beta)^wq(w)\right].\]
	Hence, the probability that a node of degree $k$ becomes infected is:
		\[1-\left(1-\mathbb{P}\right)^{k\Theta_k}\]
	where $\Theta_k$ is the probability that an edge connects a node of
	degree $k$ and an infective node. Assuming no degree correlation we
	have:
		\[\Theta_k=\frac{1}{\ave{k}}\sum_{k'} k'p(k')i_{k'}=\Theta\]
	where $i_{k'}$ is the fraction of infective nodes of degree $k'$. Now,
	translating Eq.~(\ref{SIS_mod}) in degree block approximation and
	following a similar approach to what we have done before, we obtain the
	threshold condition necessary for disease invasion:
		\begin{equation}-\frac{\ln\left(1-\mathbb{P}\right)}{\mu}>\frac{\ave{k}}{\ave{k^2}}.\label{cond_dyn}\end{equation}
	
	Below, we further explore result of Eq.~(\ref{cond_dyn}) by applying it
	to two fat-tailed probability distributions of edge-weights.
	\subsubsection{Instantaneous weight assignment, power-law distributed}\label{examples}
		Assuming that the instantaneous weight of edges follows a power-law
		distribution, $q(w)=Aw^{-\alpha}$, where $A$ is the normalizing
		constant, we could explicitly write $\mathbb{P}$ as
		\[\mathbb{P}=1-A\operatorname{Li}_{\alpha}(1-\beta),\]
		where $\operatorname{Li}_n(x)$ is the $n$th polylogarithm of $x$. Now, since:
		\[\frac{\partial\operatorname{Li}_n(x)}{\partial n}=-\sum_k {\frac{x^k}{k^n}}\ln(k)<0\]
		and
		\[\frac{\partial \operatorname{Li}_n(x)}{\partial x}=\frac{\operatorname{Li}_{n-1}(x)}{x}>0,\]
		we have that both $\partial_{\alpha}\mathbb{P}$ and
		$\partial_{\beta}\mathbb{P}$ are larger than zero, or in other
		words we have that $\mathbb{P}$ is a growing function of both of
		its variables $\beta$ and $\alpha$. Applying such knowledge to
		Eq.~(\ref{cond_dyn}) we conclude that as $\alpha$ increases, so does the
		homogeneity of instantaneous weights, while 
		the transmission probability necessary for pathogen
		persistence decreases.

	\subsubsection{Instantaneous weight assignment, negative-binomial distributed}
		Assuming that the instantaneous weight of edges follows a negative-binomial distribution, $p(w)={w+r-1\choose
		w}(1-p)^rp^w$, with $p\in(0,1)$ and $r>0$, we could write
		$\mathbb{P}$ as
		\begin{equation}\mathbb{P}=1-\left[\frac{1-p}{1-p(1-\beta)}\right]^r.\label{PPL}\end{equation}
		Now, since we are interested in exploring the behavior of the
		epidemic for decreasing heterogeneity of instantaneous weight,
		we impose $r=\frac{\ave{w}(1-p)}{p}$. Therefore,
		\[\ave{w^2}=\ave{w}\left(\ave{w}+\frac{1}{1-p}\right)\]
		or in other words, for fixed $\ave{w}$, $\ave{w^2}$ increases as
		$p$ increases. Applying the change of variable just described in
		Eq.~(\ref{PPL}) the condition for pathogen persistence,
		Eq.~(\ref{cond_dyn}), could be simplified as:
		\[-\frac{1-p}{p}\ln{\left(\frac{1-p}{1-p(1-\beta)}\right)}>\mu\frac{\ave{k}}{\ave{k^2}\ave{w}}.\]
		Now, defining $H(p,\beta)$ as the left hand of previous inequality,
		its partial derivatives are
		\[\partial_\beta H(p,\beta)=\frac{1-p}{1-p(1-\beta)}>0\]
		and
		\[
		\partial_p H(p,\beta)=\frac{1}{p^2}\left[\ln\left(\frac{1-p}{1-p(1-\beta)}\right)+\frac{\beta p}{1-p(1-\beta)}\right]<\frac{1}{p^2}\left[\left(\frac{1-p}{1-p(1-\beta)}\right)-1+\frac{\beta p}{1-p(1-\beta)}\right]<0.
		\]
		Thus, we conclude that the smaller the value of $p$, the larger the
		homogeneity of instantaneous weights, and the smaller the
		transmission probability needed for the pathogen to spread.
	
	\subsubsection{Numerical Simulations}
	It is already known, \cite{Schwarzkopf2010,Karsai2011,Perra2012,Rocha2013b}, that
	the temporal dynamic of a network reduces the probability of pathogen
	invasion. Our numerical simulations, shown on Figure~\ref{epi_curve} for
	annealed graphs confirm this result: when all weights are
	reshuffled at each time-step, $\eta=1$, the stable infected state is
	reached for larger transmission probability if compared to what happens
	on graphs with weights that remain constant with respect to time, $\eta=0$. It is also worth to
	stress that the epidemic curves obtained for $\gamma_w=1$ do not
	change comparing $\eta=0$ with $\eta=1$. This is due to the fact that
	when the weights are homogeneously distributed only little changes could
	be observable with dynamical behavior. 
	
\begin{figure}
			\begin{center}
				\includegraphics{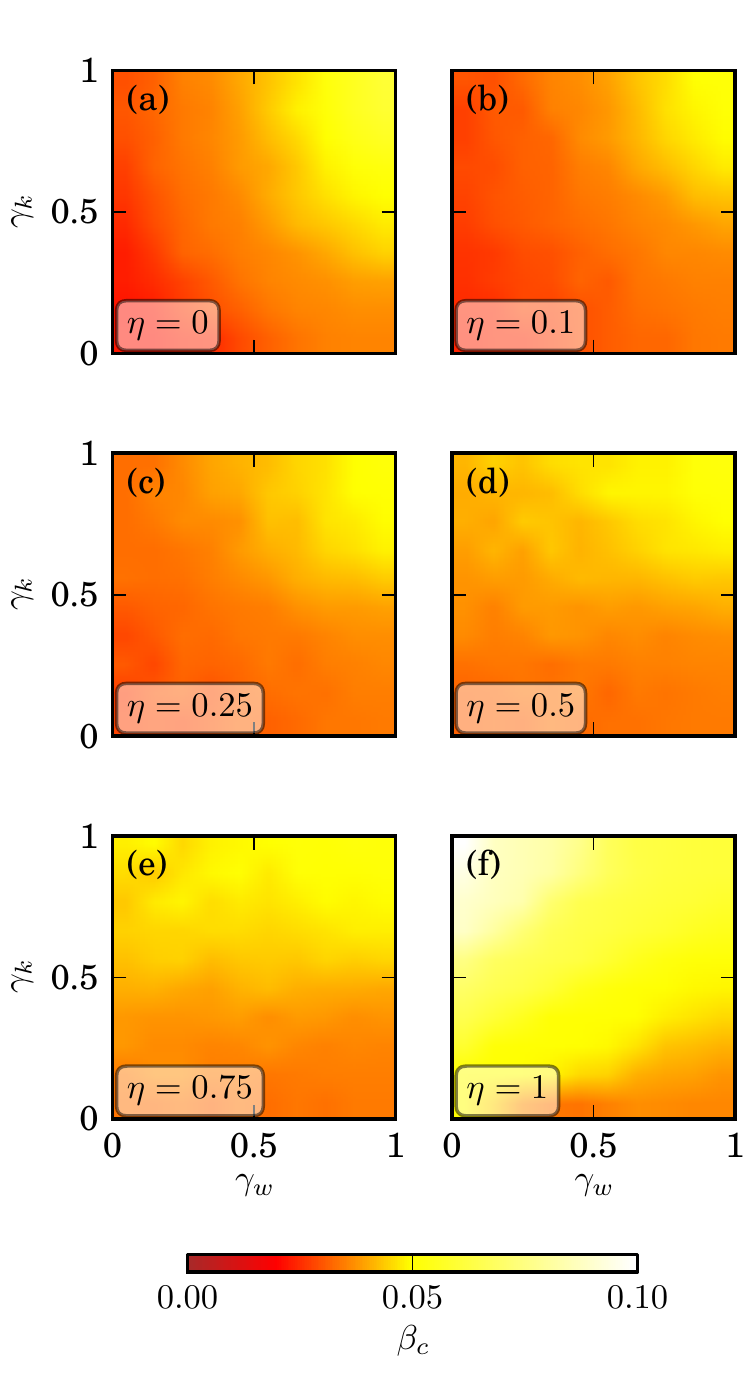}
			\end{center}
		\caption{ Critical values of transmission probability, $\beta_c$, for different fractions of dynamically assigned weights, $\eta$, in annealed graphs. Recovery probability, $\mu$, is fixed to one. }\label{mat_bc}
		\end{figure}

\begin{figure}
			\begin{center}
				\includegraphics{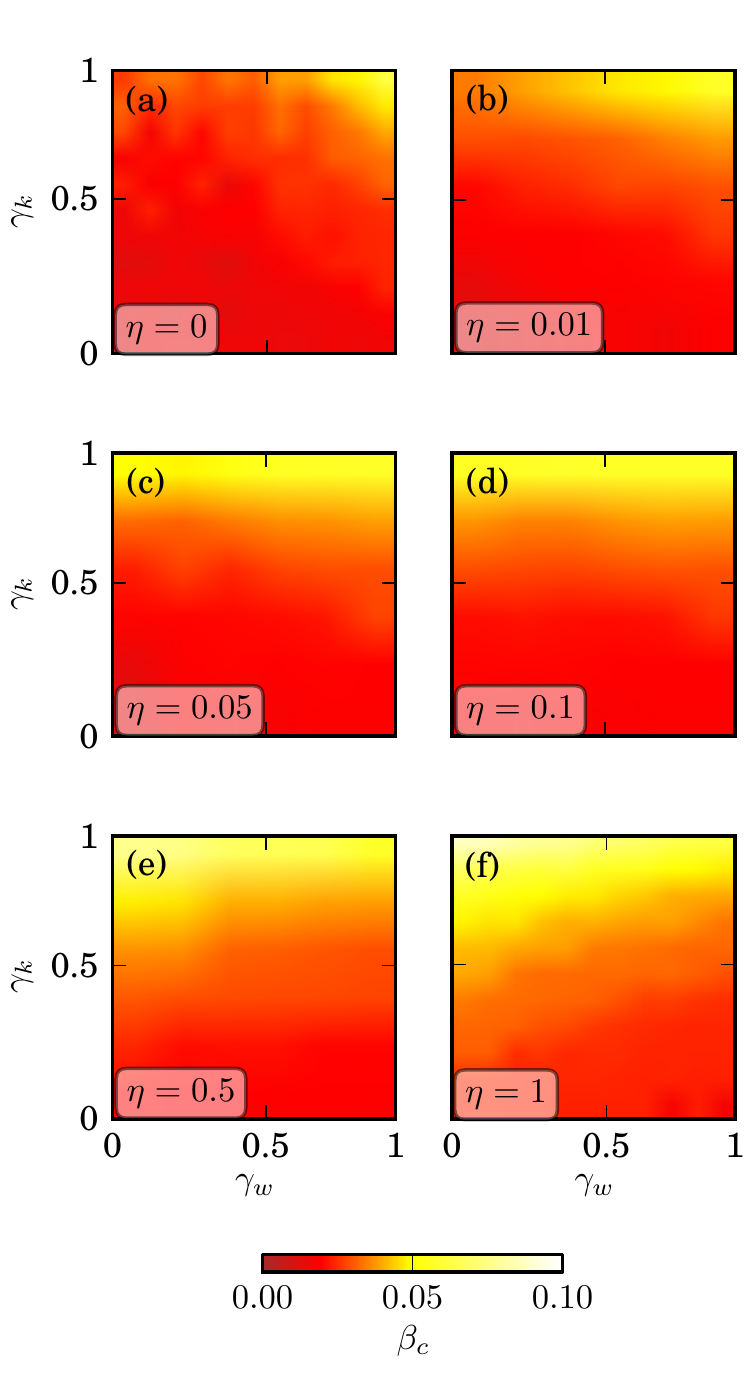}
			\end{center}
		\caption{ Critical values of transmission probability, $\beta_c$, for different fractions of dynamically assigned weights, $\eta$, in quenched graphs. Recovery probability, $\mu$, is fixed to one. }\label{QUE_mat_bc}
		\end{figure}

\begin{figure}
			\begin{center}
				\includegraphics{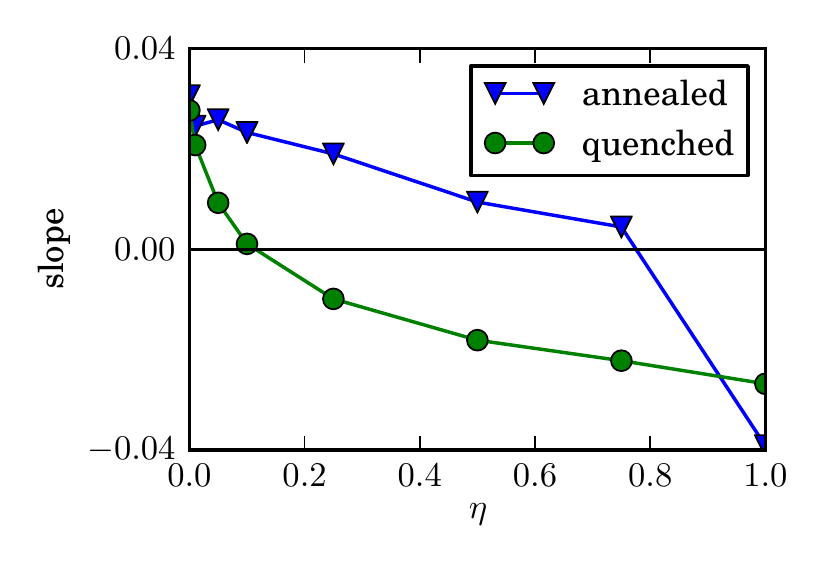}
			\end{center}
		\caption{ Slope of linear fitting of $\beta_c(\gamma_k=1,\gamma_w)$ for various $\eta$. Recovery probability, $\mu$, is fixed to one.}\label{pende}
		\end{figure}

	Consistently with the analytical results, epidemic simulations show that
	in the dynamical case the homogeneity of edge weights is a favorable
	conditions for infection spreading, conversely to what happens in the
	static case. Performing extensive numerical simulations in the whole
	$(\gamma_k, \gamma_w)$ domain, we explore how the parameter $\eta$,
	governing the dynamic reallocation of weights, influences the spreading
	dynamics. Similarly to the static case, we depict the surface $\beta_c(\gamma_k,\gamma_w)$ for
	$\eta=1$. As recovery probability we explore $\mu=0.25$ and $\mu=1$. 	
	The patterns related to a sustained disease
	transmission drastically change if compared with results obtained with
	stable weights ($\eta=0$), as shown in Figure~\ref{mat_bc_2} for
	annealed networks. Since in the two
	extreme cases, ($\eta=0,\eta=1$), the epidemic thresholds show opposite
	trends, we also explore $\eta$ by investigate some intermediate regimes
	for annealed graphs in Figure~\ref{mat_bc} (and for quenched in Figure~\ref{QUE_mat_bc}). Then,
	for each surface we linearly fit the curve
	$\beta_c(\gamma_k=1,\gamma_w)$ by a least-squares method. We plot on
	Figure~\ref{pende} the slope of the linear fit as a function of $\eta$.
	Indeed, a positive slope indicates a scenario in which the edge weight
	heterogeneity positively interacts with the infection persistence while
	a negative slope means the edge weight heterogeneity hinders disease
	spreading. Results are plotted in Figure~\ref{pende} where we recover
	the qualitative results of the previously discussed scenarios with
	$\eta=0$ and $\eta=1$ and the transient behavior between the two
	regimes. Curves obtained for quenched and annealed networks are
	different due to the different response of the two systems to changing $\eta$. In fact,
	annealed graphs are less susceptible to topology modifications, having a
	greater level of randomness if compared to quenched networks.
	Sensitivity analysis performed on $\mu$ confirms that the pattern does not change if not
	for a scale factor. This scale factor should come as no surprise, since a
	lower recovery probability makes it easier for the pathogen persist in the
	population.
		
	\section{Conclusions}	
	This work emphasizes the role of the weights and of their dynamic
	assignment to edges on the infection diffusion. As a general result we
	found that the evolving nature of weights on a graph is a limiting condition for
	the stability of infection. We suggest this possibly happens when the graph
	evolution affects the role of hub nodes in the graph. 
	
	Our results show that for weighted networks the
	pathogen spreading capability is enhanced by heterogeneity of both
	degree and weight distributions. On the contrary, this tendency is reversed when
	the weights are dynamically assigned on edges at every time-step. In
	particular, simulations show that the heterogeneity of weights acts to slow
	pathogen diffusion when we increase the fraction of edge
	weights that is dynamically assigned. Results suggest that more temporarily stable and heavier
	weights all over the network are better for pathogen spreading than some spikes of super-heavy links and a
	vast majority of very light weights where the transmission is very
	unlikely to occur. These results demonstrate the importance of dynamic
	behavior and edge weight distribution as features that should be taken
	into account when modelling infection transmissions. We strengthen our
	empirical results by analytical investigations. By using a HMF approach
	we were able to grasp the nature of the threshold $\beta_c$. Our result
	corroborates findings of previous work, \cite{Stehle2011}, indicating
	that the homogeneous assumption on the assignment of edge weights, may
	produce an entirely different behavior of epidemic spreading than
	that observed with the heterogeneous assumption.
	
	In the context of the cattle trade movements, our results could outline
	some fascinating suggestions. Let us consider the case of an infectious disease
	spreading among cattle by taking advantage of their trade movements, (e.g. the
	foot-and-mouth disease, \cite{Keeling2005}, or the diseases caused by
	bovine diarrhea virus, \cite{Tinsley2012}). It would be of interest to
	understand whether it would be a manageable and feasible containment policy
	for a state or super-state organization to intervene on commerce and
	impose a policy capable of driving the market to be more heterogeneous.
	This should decrease the probability of invasion and persistence, and it
	could be a preferable measure compared to a complete shut down of the
	trade system. Extensive research in this direction will be subject
	of future research.
	
	{Future work will be devoted to the analytical exploration of spreading on quenched networks.} Working towards this goal, it might be useful to extend this
	framework to some real-world epidemic scenarios, also taking into account
	more complex transmission routes than SIS. An important
	extension to our research would be to integrate into our framework degree,
	weights and temporal correlations as observed in real-world systems. \\
	\begin{acknowledgments}
		Authors thank Bryan N. Iotti for fruitful discussions. LF acknowledges support from the Lagrange Project, CRT and ISI
		Foundation. PB and MG acknowledges Compagnia di San Paolo and MG
		acknowledges local research funding of the University of Torino. 
	\end{acknowledgments}
	\bibliography{biblio}
\end{document}